\documentclass[preprint,12pt]{elsarticle}

\usepackage{amssymb}

\usepackage{lineno}

\begin{document}
%\begin{linenumbers}
\begin{frontmatter}



\title{On the Covariance of the Charge Form Factor in the Transition Radiation Energy Spectrum of a Beam at Normal Incidence onto a Metallic Screen}

\author{Gian Luca Orlandi}
\ead{gianluca.orlandi@psi.ch}
\address{Paul Scherrer Institut, 5232 Villigen PSI, CH.}

\begin{abstract}
A charge-density-like covariance is expected to characterize the
transition radiation energy spectrum of a $N$ electron bunch as far
as the charge form factor is intended to account for bunch-density
effects in the radiation emission. The beam charge passing from a
single electron to a high density electron bunch, the covariance of
the transition radiation energy is expected to evolve from a
charge-point-like to a charge-density-like one. Besides covariance,
the radiation energy spectrum is expected to conform to the temporal
causality principle: the $N$ single electron amplitudes composing
the radiation field are expected to propagate from the metallic
screen with relative emission phases causally correlated with the
temporal sequence of the $N$ particle collisions onto the metallic
screen. In the present paper, the case of a $N$ electron bunch
hitting at a normal angle of incidence a flat metallic surface with
arbitrary size and shape will be considered. For such an
experimental situation, the distribution function of the $N$
electron longitudinal coordinates rules the temporal causality
constraint into the transition radiation energy spectrum. The
covariance feature of the transition radiation energy spectrum deals
instead with the Lorentz invariance of the projection of the $N$
electron spatial density in the transverse plane with respect to the
direction of motion of the $N$ electron beam. Because of the
invariance of the $N$ electron transverse density under a Lorentz
transformation with respect to the direction of motion of the
electron beam, the $N$ single electron radiation amplitudes
composing the radiation field show a covariant dependence on the
distribution function of the $N$ electron transverse coordinates,
the relative emission phases of the $N$ single electron radiation
amplitudes being indeed only a function of the $N$ electron
longitudinal coordinates because of the temporal causality
constraint. As a consequence of the temporal causality and the
covariance, both the temporal coherent and incoherent components of
the radiation energy spectrum bear the covariant imprinting of the
distribution function of the $N$ electron transverse coordinates as
in the following argued.
\end{abstract}

\end{frontmatter}

\linenumbers

\section{Introduction}

A relativistic charge in a rectilinear and uniform motion can
originate an electromagnetic radiative mechanism crossing the
dielectric interface between two different media. The fast dipolar
oscillation of the polarization charge, induced on the dielectric
interface by the relativistic charge, generates indeed an
instantaneous, broad spectral band, radially polarized and highly
directional radiation emission, the so called transition radiation
\cite{gifr,gari,gari2,frank2,bass,frank,ter,ginz}. A photon pulse
propagates backward and forward from the dielectric interface
according to a double conical spatial distribution whose angular
aperture scales down with the Lorentz $\gamma$ factor of the charge.
The higher and steeper the discontinuity of the dielectric constant
across the interface, the more intense the radiation emission.

Thanks to the relativistic origin, transition radiation can be
fruitfully exploited for energy detection or mass identification of
high energy particles
\cite{prun,alik,yuan,alik2,wart2,cherry,fabj,chu,pies,fior,wakely,andro}.
Transition radiation based diagnostics is commonly used in a
particle accelerator to monitor the transverse
\cite{wart,boss,rule,lump,rule3,loos,denar,catra,leb,artru,cast2,poty,shulga,bravin,sakamoto,welch,holloway}
or the longitudinal profile of a charged beam
\cite{happek,shibata,barry2,taka,kung,shib2,lai,rosen,lihn,schlott,schnei,watana,geitz,mihal,tilborg,sutter,marsch,casalbuoni}.
Transition radiators, typically made of a thin metallic foil or a
thin polished Aluminium coating on dielectric substrates, respond as
an ideal conductor over a wide spectral range, from the long
wavelength to far beyond the visible optical region \cite{hand}.
This experimental situation is supposed in the present work. In
particular, the case of a $N$ electron bunch colliding at a normal
angle of incidence onto a flat ideal conductor surface will be
considered in the following.

An ideal conductor surface being assimilable to a double layer of
charge, transition radiation emission can be thus schematized as
the result of the interaction of an incident charge with the
conduction electrons of the ideal conductor surface. As the
relativistic charge approaches the metallic surface, the charge
induced conduction electrons freely move in the transverse plane
to maintain the metallic surface equipotential. The sudden and
fast dipolar oscillation of the double layer of charge is thus
responsible for the emission of the radiation pulse as confirmed
by the emission of backward radiation.

In the relativistic limit, the electromagnetic field travelling with
a relativistic electron can be assimilated to a transverse
electromagnetic wave, the so called {\it virtual quanta} field
\cite{ter,jack}. In the case of a single relativistic electron, the
harmonic component of the electric field at a given wavelength
$\lambda$ extends indeed in the transverse plane over an efficacious
range in the order of $\gamma\lambda/2\pi$, the ratio of the
longitudinal to the transverse component being $1/\gamma^2$. The
{\it virtual quanta} field showing the nature of a quasi-plane wave
front, the radiation emission can be formally described as the
result of the wave propagation of the {\it virtual quanta} scattered
by the metallic surface according to the Huygens-Fresnel principle
\cite{born}. Under the far field approximation \cite{ter}, the
harmonic components of the transition radiation field at the
observation point can be calculated as the Fourier transform of the
{\it virtual quanta} field with respect to the spatial coordinates
of the radiator surface. The transition radiation energy spectrum
can be finally obtained as the flux of the Poynting vector.

A single and individually radiating electron and a high density
electron beam are in general different from the point of view of the
features of the spectral and the angular distribution of the emitted
radiation. Compared to the case of a single and individually
radiating electron, bunch-density effects may strongly affect the
electromagnetic radiative mechanism by an electron beam. For a fixed
energy of the beam, the number and the angular distribution of the
photons radiated at a given wavelength may indeed change as a
function of the density of the electron beam. In the formula of the
radiation energy spectrum, the charge form factor accounts for bunch
collective effects. This is, in principle, defined as the square
module of the Fourier transform of the distribution function of the
particle density of the charged beam.

Aim of the present paper is to verify how well such a theoretical
definition of the charge form factor fits into a covariant
formulation of the radiation energy spectrum or if, instead, under
the covariance and the temporal causality constraints, the radiation
energy spectrum shows a dependence on the charge density that goes
beyond the above mentioned formal definition of the charge form
factor. In fact, the transverse density of the electron beam is a
relativistic invariant under a Lorentz transformation with respect
to the direction of motion of the beam. The invariance of the
projection of the electron beam density onto the transverse plane
with respect to the direction of motion is expected to leave, on the
{\it virtual quanta} field and, consequently, on the radiation
field, a covariant mark whose observability, for a given harmonic
component of the radiation field, only depends on the transverse
dimension of the electron beam compared to the transverse component
of the wave-vector which is itself a Lorentz invariant with respect
to the direction of motion of the beam. In the following, the
charge-density-like covariance of the transition radiation energy
spectrum by an electron beam will be investigated. First, the formal
steps leading from the electromagnetic field of a $N$ electron bunch
to the transition radiation energy spectrum will be analyzed with
regard to the temporal causality constraint. Finally, the covariance
of the given formula of the electromagnetic field of the $N$
electron bunch will be checked by verifying that, under a Lorentz
transformation from the laboratory to the rest reference frame, the
expected formula of the electric field of a charge distribution at
rest is obtained.

\section{Virtual quanta field and transition radiation energy spectrum of a $N$ electron beam}\label{pseudo-virtual}

In the following, $N$ electrons in motion with a common rectilinear
and uniform velocity $\vec{w}$ along the $z$ axis of the laboratory
reference frame are supposed to strike, at a normal angle of
incidence, onto a flat ideal conductor surface $S$. The radiator,
being in vacuum and placed in the plane $z=0$ of the laboratory
reference frame, is supposed to have an arbitrary size and shape.
Compared to the synchrotron motion of a single electron in the
bunch, the time scale of the emission of the radiation pulse by the
entire electron bunch can be considered instantaneous. Consequently,
in modeling the radiation emission from the electron bunch,
collisions between electrons in the bunch can be neglected and the
$N$ electron bunch can be thus described in terms of a
``frozen-in-time" spatial distribution function, i.e., a spatial
distribution function that is invariant under a time-space
translation in the laboratory reference frame (see also \ref{cov}).
With reference to such an experimental context, the temporal
sequence of the $N$ electron collisions onto the metallic screen is
only ruled by the distribution function of the $N$ electron
longitudinal coordinates $z_{0j}$ ($j=1,..,N$) which, for the sake
of ease, are taken at the time $t=0$ when the center of mass of the
electron bunch is supposed to strike the metallic surface.

On the radiator surface $S$ ($z=0$), the harmonic components of the
transverse electric field of both the relativistic charge
($E_{x,y}^{vq}$) and the charge induced conduction electrons
($E_{x,y}$) satisfy the following boundary constraint:
\begin{equation}
E_{x,y}^{vq}(x,y,z=0,\omega)+E_{x,y}(x,y,z=0,\omega)=0.\label{uno}
\end{equation}
The boundary condition above rules the dipolar oscillation of the
the conduction electrons, which is induced on the ideal conductor
surface by the incident relativistic charge, and explains why a
radiation field needs to fly away from the boundary surface in order
to maintain it equipotential. The transition radiation emission can
be thus interpreted as the result of the scattering of an
electromagnetic wave from a metallic surface and described, on the
basis of the Huygens-Fresnel principle, as the wave propagation of
the transverse component ($E_{x,y}^{vq}$) of the {\it virtual
quanta} field \cite{ter,jack} from the surface $S$ ($z=0$) to the
observation point. The transition radiation field ($E_{x,y}^{tr}$)
can be calculated by means of the Helmholtz-Kirchhoff integral
theorem \cite{jack,born}. Under the far-field approximation, this
integral can be finally expressed as the Fourier transform of the
harmonic component of the {\it virtual quanta} field with respect to
the spatial coordinates $\vec{\rho}=(x,y)$ of the radiator surface
$S$ \cite{ter,jack,born}:
\begin{equation}
E_{x,y}^{tr}(\vec{\kappa},\omega)=\frac{k}{2\pi R} \int\limits_S
d\vec{\rho}\,E_{x,y}^{vq}(\vec{\rho},\omega)e^{-i\vec{\kappa}\cdot\vec{\rho}}\label{due}
\end{equation}
where $k=\omega/c$ is the wave-number, $\vec{\kappa}=(k_x,k_y)$ is
the transverse component of the wave-vector and $R$ is the distance
from the radiator surface to the observation point that, in the
present context, is supposed to be on the $z$-axis of the laboratory
reference frame.

In the case of $N$ electrons in a rectilinear and uniform motion
with a common velocity $\vec{w}=(0,0,w)$, the harmonic component of
the {\it virtual quanta} electric field at the radiator surface $S$
($z=0$) reads, see
\cite{gian,gianbis,gianter,gianquater,gian5,gian6} and also
\ref{vq-lab}:
\begin{eqnarray}
E_{x,y}^{vq}(x,y,z=0,\omega)=\frac{i\,e}{w\pi}\sum_{j=1}^{N}e^{-i(\omega/w)z_{0j}}\int
d\vec{\tau}
\,e^{i\vec{\tau}\cdot\vec{\rho}}\,\frac{\tau_{x,y}\,e^{-i\vec{\tau}\cdot\vec{\rho}_{0j}}}{\tau^2+\alpha^2},\label{prima-eq}
\end{eqnarray}
where $\alpha=\frac{\omega}{w\gamma}$, while
$\vec{\rho}_{0j}=(x_{0j},y_{0j})$ and $z_{0j}$ $(j=1,..,N)$ are,
respectively, the transverse and the longitudinal coordinates of the
electrons in the laboratory reference frame at the time $t=0$ when
the center of mass of the electron bunch is supposed to strike the
metallic surface. Taking into account
Eqs.(\ref{due},\ref{prima-eq}), the transition radiation electric
field by a $N$ electron bunch reads:
\begin{eqnarray}
E_{x,y}^{tr}(\vec{\kappa},\omega)=
\sum_{j=1}^{N}H_{x,y}(\vec{\kappa},\omega,\vec{\rho}_{0j})\,e^{-i(\omega/w)z_{0j}}\label{tre}
\end{eqnarray}
where
\begin{eqnarray}
H_{\mu}(\vec{\kappa},\omega,\vec{\rho}_{0j})=H_{\mu,j}=\frac{iek}{2\pi^2Rw}\int\limits_S
d\vec{\rho}\int d\vec{\tau}
\frac{\tau_{\mu}e^{-i\vec{\tau}\cdot\vec{\rho}_{0j}}}{\tau^2+\alpha^2}
e^{i(\vec{\tau}-\vec{\kappa})\cdot\vec{\rho}}\label{quattro1}
\end{eqnarray}
with $\mu=x,y$.

With reference to Eqs.(\ref{tre},\ref{quattro1}), the transition
radiation energy spectrum by a $N$ electron beam can be finally
calculated as the flux of the Poynting vector:
\begin{eqnarray}
\frac{d^2I}{d\Omega
d\omega}&&=\frac{cR^2}{4\pi^2}\left(\left|E_{x}^{tr}(k_x,k_y,\omega)
\right|^2+\left|E_{y}^{tr}(k_x,k_y,\omega)\right|^2\right)=\label{cento-uno}\\
\nonumber
&&=\frac{cR^2}{4\pi^2}\sum_{\mu=x,y}\left(\sum_{j=1}^{N}\left|H_{\mu,j}
\right|^2+\sum_{j,l(j\neq l)=1}^{N}e^{-i(\omega/w)(z_{0j}-z_{0l})}
H_{\mu,j}H^*_{\mu,l}\right)
\end{eqnarray}
where in previous equation the radiator surface $S$ has, in general,
an arbitrary shape and size (either infinite $S=\infty$ or finite
$S<\infty$).

The {\it virtual quanta} field - see Eq.(\ref{prima-eq}) - shows a
field structure as a train of $N$ travelling transverse waves
hitting the metallic surface with a relative phase delay only
dependent on the difference between the longitudinal coordinates of
the $N$ electrons $z_{0j}$ $(j=1,..,N)$. In the same way, the
radiation field - Eqs.(\ref{tre},\ref{quattro1}) - results from the
linear addition of $N$ single particle field amplitudes $H_{\mu,j}$
whose relative emission phases - $e^{-i(\omega/w)z_{0j}}$,
$(j=1,..,N)$ - from the metallic surface causally depend on the
temporal sequence of the $N$ electron collisions onto the metallic
screen.

The temporal causal structure characterizing the transition
radiation field of the $N$ electron bunch -
Eqs.(\ref{tre},\ref{quattro1}) - is also a feature of the radiation
energy spectrum - see Eq.(\ref{cento-uno}) - where the $N$ single
particle field amplitudes $H_{\mu,j}$ interfere indeed each other
via a relative phase factor only dependent on the relative
difference of the electron longitudinal coordinates $z_{0j}$
$(j=1,..,N)$. In fact, $(1)$ the radiator surface $S$ being in
principle arbitrary and $(2)$ the formal expression of the
transition radiation energy spectrum being expected to be invariant
whether the radiator surface $S$ is finite ($S<\infty$) or infinite
($S=\infty$) and whatever is its shape, the $N$ single electron
radiation field amplitudes $H_{\mu,j}$ $(j=1,..,N)$ -
Eqs.(\ref{tre},\ref{quattro1}) - can be formulated in terms of an
implicit integral form and treated as a sort of special function
whose numerical value can be only formally and implicitly stated in
view of the general formulation of the radiation energy spectrum,
Eq.(\ref{cento-uno}). Therefore, thanks to the arbitrariness of the
radiator surface $S$ and to the consequent implicit integral
formulation of the $N$ single electron radiation field amplitudes,
the formula of the transition radiation energy spectrum -
Eq.(\ref{cento-uno}) - explicitly shows the temporal causal feature:
the interference between the different single electron field
amplitudes $H_{\mu,j}$ is indeed ruled by the distribution function
of the electron longitudinal coordinates $z_{0j}$ $(j=1,..,N)$ via
the phase factor $e^{-i(\omega/w)(z_{0j}-z_{0l})}$ $(l \neq j)$.

About the electron transverse coordinates
$\vec{\rho}_{0j}=(x_{0j},y_{0j})$ $(j=1,..,N)$, they play the role
to contribute, as a function of the electron displacement from the
$z$-axis where the radiation field is supposed to be observed, not
only to the global phase factor of the single electron field
amplitude $H_{\mu,j}$ $(j=1,..,N)$ at the observation point but also
to the single electron field amplitude itself. It can be indeed
demonstrated \cite{gian-finite} that, in the case of a round screen
with a finite radius, the integral representing the $N$ single
electron amplitudes composing the radiation field -
Eqs.(\ref{tre},\ref{quattro1}) - can be solved showing the structure
of the product of a real amplitude and a phase factor both depending
on the electron transverse coordinates:
\begin{eqnarray}
H_{\mu,j}=H_{\mu}(\vec{\kappa},\omega,\vec{\rho}_{0j})=
A_{\mu}(\vec{\kappa},\omega,\vec{\rho}_{0j})e^{-i\vec{\kappa}\cdot\vec{\rho}_{0j}}.\label{cov-end-milleeuno}
\end{eqnarray}
On the basis of Eq.(\ref{cov-end-milleeuno}), the formula of the
radiation energy spectrum given in Eq.(\ref{cento-uno}) shows the
well known dependence on the three-dimensional form factor as well
as an intrinsic dependence of the $N$ single electron radiation
field amplitudes on the electron transverse coordinates. The
transverse density of the $N$ electrons is indeed an invariant under
a Lorentz transformation with respect to the direction of motion of
the beam. The transverse coordinates of the $N$ electrons are thus
expected: ($1$) to contribute to the relative phase delay of the $N$
electron field amplitudes at the observation point as a function of
the transverse displacement of the $N$ electrons with respect to the
beam axis where the radiation field is supposed to be observed;
($2$) because of the Lorentz invariance of the transverse projection
of the $N$ electron density, to leave a covariant mark on the $N$
electron radiation field amplitudes - Eqs.(\ref{tre},\ref{quattro1})
- and, consequently, on both the temporal coherent and incoherent
components of the radiation energy spectrum, Eq.(\ref{cento-uno}).

In conclusion, in the most general case of an electron bunch at a
normal angle of incidence onto a radiator surface $S$ with an
arbitrary size and shape (finite $S<\infty$ or infinite
$S=\infty$), Eq.(\ref{cento-uno}) represents a temporal causal and
covariance consistent formulation of the transition radiation
energy spectrum.

\section{Covariance of the charge form factor in the transition radiation energy
spectrum}\label{capitolo-cov-ff}

\subsection{Covariance of the electron bunch density}\label{parag-cov-bunch-density}

The covariance of the charge form factor and, in general, of the
transition radiation energy spectrum - as formulated in
Eqs.(\ref{tre},\ref{quattro1},\ref{cento-uno}) - will be analyzed in
the following paragraph. For such a purpose, the Fourier
transformations of fields and charge distributions as normally
defined in the ordinary $3$-dimensional space will be reformulated
in an explicitly covariant form as $4$-dimensional Fourier
representations in the conjugate Fourier wavevector-frequency
$4$-space $(k_x,k_y,k_z,\omega/c)$ of the space-time $(x,y,z,ct)$ in
the laboratory reference frame $\mathbb{R}$ or in the conjugate
Fourier $4$-space $(k'_x,k'_y,k'_z,\omega'/c)$ of the space-time
$4$-vector $(x',y',z',ct')$ in the rest reference frame
$\mathbb{R'}$, see \ref{lorentz}, \ref{cov}, \ref{cov-rest} and
\ref{vq-lab}. The case of the covariant transformation of the
spatial density of a $N$ electron bunch, as in the following
described, will exemplify the above mentioned explicitly covariant
extension of a Fourier representation of a field or a charge
distribution from the ordinary $3$-dimensional space to the Lorentz
$4$-space.

In the present context, the $N$ electrons of the bunch are described
by a ``frozen" in time spatial distribution function: all the
electrons are supposed to move with a rectilinear and uniform
velocity $\vec{w}=(0,0,w)$ along the $z$ axis of the laboratory
reference frame $\mathbb{R}$. In the reference frame $\mathbb{R'}$,
where the $N$ electrons are at rest, the distribution function of
the spatial density of the electron bunch reads:
\begin{equation}
\rho'(\vec{r'},t')=\rho'(\vec{r'},0)=\sum_{j=1}^N\delta(\vec{r'}-
\vec{r'}_j),\label{quattro}
\end{equation}
where $\vec{r'}_j=(x'_j,y'_j,z'_j)$ $(j=1,..,N)$ are the spatial
coordinates of the $N$ electrons at the time $t'$ in $\mathbb{R'}$.
In the ordinary $3$-space $\vec{r'}=(x',y',z')$ and the conjugate
Fourier $3$-space $\vec{k'}=(k'_x,k'_y,k'_z)$ of $\mathbb{R'}$, the
spatial density of the $N$ electrons - Eq.(\ref{quattro}) - reads
\begin{equation}
\rho'(\vec{r'},t')=\frac{1}{(2\pi)^3}\sum_{j=1}^{N}\int
d\vec{k'}e^{i\vec{k'}\cdot(\vec{r'}-\vec{r'_j})}.\label{cinque}
\end{equation}
The above Fourier transformation of the spatial density of the $N$
electron bunch is defined in the space $(x',y',z')$ and in the
conjugate Fourier space $(k'_x,k'_y,k'_z)$ of the rest frame of
reference $\mathbb{R'}$, which are subspaces of the space-time
$(x',y',z',ct')$ and of the wavevector-frequency
$(k'_x,k'_y,k'_z,\omega'/c)$, respectively. Consequently, the charge
density, as expressed by Eq.(\ref{cinque}) in the ordinary
$3$-dimensional space of $\mathbb{R'}$, cannot explicitly show the
expected covariance under a Lorentz transformation from the rest to
the laboratory reference frame
$(\mathbb{R'}\Longrightarrow\mathbb{R})$. An explicitly covariant
Fourier representation of the spatial density of the $N$ electrons
can be obtained by extending the Fourier transformation as given in
Eq.(\ref{cinque}) into the Fourier conjugate $4$-space
$(k'_x,k'_y,k'_z,\omega'/c)$ of the reference frame $\mathbb{R'}$,
for details see \ref{cov}. With reference to \ref{cov}, the spatial
density of the $N$ electron bunch - as given in Eq.(\ref{cinque}) -
can be represented in an explicitly covariant form in the space-time
and wavevector-frequency $4$-spaces of the rest reference frame
$\mathbb{R'}$ as follows:
\begin{eqnarray}
\rho'(\vec{r'},t')=\frac{c}{(2\pi)^4}\int
d^4\kappa'e^{i\kappa'\cdot\xi'}\rho(\vec{k'},\omega')\label{sei}
\end{eqnarray}
where $d^4\kappa'=d\vec{k}'d(\omega'/c)$ and
\begin{equation}
\rho(\vec{k'},\omega')=2\pi\left(\sum_{j=1}^{N}e^{-i\kappa'\cdot\xi'_j}\right)\delta(\omega'),\label{sette}
\end{equation}
where $\xi'_j=(x'_j,y'_j,z'_j,ct')$ $(j=1,..,N)$ are the space-time
$4$-vectors of the electron coordinates, while $\xi'=(x',y',z',ct')$
and $\kappa'=(k'_x,k'_y,k'_z,\omega'/c)$ are the space-time and the
conjugate Fourier wavevector-frequency 4-vectors in the rest
reference frame $\mathbb{R'}$, respectively.

The covariance of the Fourier representation of the spatial density
of $N$ electron bunch - as represented in Eq.(\ref{cinque}) or,
equivalently, in Eqs.(\ref{sei},\ref{sette}) - can be now easily
checked, see also \ref{cov}. With reference to
Eqs.(\ref{sei},\ref{sette}), $\rho(\vec{k'},\omega')$ is the only
non-Lorentz-invariant quantity in the integrand of Eq.(\ref{sei}).
Moreover, according to Eqs.(\ref{sette}), the covariance of
$\rho(\vec{k'},\omega')$ only depends on how the delta Dirac
function $\delta(\omega')$ transforms under a Lorentz
transformation. Under a Lorentz transformation from the rest to the
laboratory reference frame $(\mathbb{R'}\Longrightarrow\mathbb{R})$
- see \ref{lorentz} and \ref{cov} - the delta Dirac function
transforms indeed as
\begin{equation}
\delta(\omega')=\delta[\gamma(\omega-wk_z)]=\frac{\delta(\omega-\vec{w}\cdot\vec{k})}{\gamma}.\label{sette-bis}
\end{equation}
In conclusion, under a Lorentz transformation
$(\mathbb{R'}\Longrightarrow\mathbb{R})$, the $N$ electron spatial
density - as represented by Eq.(\ref{cinque}) in the ordinary
$3$-space or, equivalently, by Eqs.(\ref{sei},\ref{sette}) in the
Lorentz $4$-space - transforms in agreement with the expected
time-like covariance as
\begin{equation}
\rho'(\vec{k'},\omega')=\frac{1}{\gamma}\rho(\vec{k},\omega),\label{otto}
\end{equation}
where
\begin{eqnarray}
\rho(\vec{k},\omega)=2\pi\left(\sum_{j=1}^{N}e^{-i\kappa\cdot\xi_j}\right)\delta(\omega-\vec{w}\cdot\vec{k})
\label{sei-bis}
\end{eqnarray}
is the $4$-dimensional Fourier transformation of the $N$ electron
spatial density in the laboratory reference frame $\mathbb{R}$, see
also \ref{cov}.

\subsection{Covariance of the electron bunch form factor}

In the present context, a bunch of $N$ electrons moving with a
common rectilinear and uniform velocity $\vec{w}=(0,0,w)$ along the
$z$-axis of the laboratory reference frame is supposed to strike, at
a normal angle of incidence, a flat metallic screen placed in the
plane $z=0$. For such an experimental situation, Equations
(\ref{tre},\ref{quattro1}) and (\ref{cento-uno}) formulate the
transition radiation energy spectrum in the most general case of a
radiator surface $S$ having an arbitrary shape and size (either
finite $S<\infty$ or infinite $S=\infty$).

This formulation of the radiation energy spectrum is very general.
It is indeed expressed in terms of the original discrete
distribution function of the $N$ electron coordinates: no continuous
limit approximation is indeed applied to the distribution function
of the electron spatial coordinates. Furthermore, the size and the
shape of the radiator surface being indeed arbitrary,
Eqs.(\ref{tre},\ref{quattro1},\ref{cento-uno}) represent a very
general and implicit formulation of the transition radiation energy
spectrum of a $N$ electron bunch which is invariant in relation to
the size and shape of the radiator.

This formulation of the radiation energy spectrum also meets the
temporal causality constraint. For the electromagnetic radiative
mechanism of the $N$ electron bunch that is described in the present
paper, the causality constraint rules the temporal correlation
between the single electron collision onto the metallic screen and
the consequent emission from the metallic surface of the
corresponding single electron radiation field amplitude $H_{\mu,j}$
$(j=1,..,N)$, see Eqs.(\ref{tre},\ref{quattro1}). In the considered
experimental context, the temporal sequence of the $N$ electron
collisions onto the metallic screen is only ruled by the
distribution function of the longitudinal coordinates $z_{0j}$
$(j=1,..,N)$ of the $N$ electrons. Thanks to the implicit integral
formulation of the $N$ single electron amplitudes $H_{\mu,j}$
$(j=1,..,N)$ composing the radiation field - see
Eqs.(\ref{tre},\ref{quattro1}) - the general formula of the
transition radiation energy spectrum - Eq.(\ref{cento-uno}) -
explicitly meet the temporal causality constraint. In
Eq.(\ref{cento-uno}) indeed, the dependence of the $N$ single
electron field amplitudes on the electron transverse coordinates
$\vec{\rho}_{0j}=(x_{0j},y_{0j})$ $(j=1,..,N)$ being encoded in the
implicit integral formulation given in
Eqs.(\ref{tre},\ref{quattro1}), the interference between the $N$
single electron field amplitudes $H_{\mu,j}$ $(j=1,..,N)$ is a
direct function of the corresponding emission phases from the
metallic surface, $e^{-i(\omega/w)(z_{0j}-z_{0l})}$ $(l \neq j)$, in
agreement with the temporal causal principle.

About the covariance of the transition radiation energy spectrum of
the $N$ electron bunch as formulated in
Eqs.(\ref{tre},\ref{quattro1},\ref{cento-uno}), this can be argued
on the basis of the following analysis: first, verify the covariance
of the formal steps leading from the {\it virtual quanta} field of
the $N$ electron bunch to the resultant radiation energy spectrum,
see Eq.(\ref{prima-eq}) and
Eqs.(\ref{tre},\ref{quattro1},\ref{cento-uno}); finally, verify the
covariance of the given expression of the {\it virtual quanta} field
under a Lorentz transformation from the laboratory reference frame
$\mathbb{R}$ to the reference frame $\mathbb{R'}$ of rest of the $N$
electrons.

About the covariance of the formal procedure leading from the {\it
virtual quanta} field to the radiation field, the following
observation can be done. On the basis of the far-field
implementation of the Helmholtz-Kirchhoff integral theorem
\cite{ter,jack,born}, a given harmonic component of the transition
radiation field of the $N$ electron bunch can be obtained by Fourier
transforming the corresponding {\it virtual quanta} field with
respect to the spatial coordinates of the radiator surface $S$, see
Eq.(\ref{due}). The calculation of such a Fourier transformation
only involves the coordinates $(x,y)$ of the radiator surface $S$
and the related conjugate Fourier coordinates $(k_x,k_y)$ which are
transverse with respect to the direction of motion of the electron
bunch and thus invariant under a Lorentz transformation. Therefore,
the covariance properties of the {\it virtual quanta} field are
entirely and unalterably transferred into the transition radiation
field - see Eq.(\ref{prima-eq}) and Eq.(\ref{tre}) - via the Fourier
transformation as defined in Eq.(\ref{due}).

Finally, about the covariance of the charge form factor formulation
in the transition radiation energy spectrum -
Eqs.(\ref{tre},\ref{quattro1},\ref{cento-uno}) - this can be
verified by analyzing how the {\it virtual quanta} field of the $N$
electron bunch transforms under a Lorentz transformation from the
laboratory to the rest reference frame
($\mathbb{R}\Longrightarrow\mathbb{R'}$), see also \ref{lorentz}.

In the present context, a ``frozen-in-time" electron bunch is
considered, i.e., the $N$ electrons in motion with the same velocity
$\vec{w}=(0,0,w)$ are described by a distribution function of the
spatial coordinates that is invariant under a time-space translation
in the laboratory reference frame $\mathbb{R}$, see also \ref{cov}
\begin{displaymath}
\left\{
\begin{array}{l}
\vec{r}=\vec{r}_0+\vec{w}t\\
\vec{r}_j=\vec{r}_{0j}+\vec{w}t
 \end{array} \right.
\end{displaymath}
where $\vec{r}_{0j}$ $(j=1,..,N)$ are the $N$ electron spatial
coordinates at the time $t=0$ when, for instance, the center of mass
of the $N$ electron coordinates is supposed to cross the plane
$z=0$.

For such an experimental situation, with reference to
\cite{gian,gianbis,gianter,gianquater,gian5,gian6}, the expression
of the charge electric field reads, see also \ref{vq-lab}:
\begin{eqnarray}
\vec{E}(\vec{k},\omega)=-i(8\pi^2e)\frac{[\vec{k}-(\omega\vec{w}/c^2)]}{[k^2-(\omega/c)^2]}
\left(\sum_{j=1}^{N}e^{-i\vec{k}\cdot\vec{r}_{0j}}\right)\delta(\omega-\vec{w}\cdot\vec{k}).\label{cov-bisbis}
\end{eqnarray}
The expression above of the electric field of the $N$ electron bunch
does not explicitly show the expected covariance as the argument of
the phase factor in Eq.(\ref{cov-bisbis}) clearly indicates.
Following the procedure already described in previous subsection
\ref{parag-cov-bunch-density} and in \ref{cov}, \ref{cov-rest} and
\ref{vq-lab}, the explicit covariance of the electric field of the
$N$ electrons can be retrieved by extending the Fourier
representation of the charge electric field from the ordinary
$3$-dimensional spaces of the spatial coordinates $\vec{r}=(x,y,z)$
and conjugate wave-vector coordinates $\vec{k}=(k_x,k_y,k_z)$ into
the Lorentz $4$-space of the space-time $\xi=(\vec{r},ct)$ and the
conjugate Fourier $4$-space of the wavevector-frequency
$\kappa=(\vec{k},\omega/c)$. With reference to the time-space
translation of the spatial coordinates of the $N$ electrons in the
laboratory reference frame $\mathbb{R}$, see Equations above, the
Fourier representation of the electric field of the $N$ electrons -
given in Eq.(\ref{cov-bisbis}) - can be reformulated in the
following explicitly covariant form, see also
Eqs.(\ref{vq-lab-uno},\ref{vq-lab-due}) in \ref{vq-lab}:
\begin{eqnarray}
\vec{E}(\vec{r},t)=\frac{1}{(2\pi)^4}\int d\vec{k}d\omega \,
e^{i(\vec{k}\cdot\vec{r}-\omega
t)}\vec{E}(\vec{k},\omega)=\frac{c}{(2\pi)^4}\int d^4\kappa \,
e^{i\kappa\cdot\xi}\vec{E}(\vec{k},\omega)\label{cov-uno}
\end{eqnarray}
where $d^4\kappa=d\vec{k}\,d(\omega/c)$ and, taking into account the
following time-space translation $\vec{r}_j=\vec{r}_{0j}+\vec{w}t$
(see coordinates transformation above),
\begin{eqnarray}
\vec{E}(\vec{k},\omega)&&=-i(8\pi^2e)\frac{[\vec{k}-(\omega\vec{w}/c^2)]}{[k^2-(\omega/c)^2]}
\left(\sum_{j=1}^{N}e^{-i\vec{k}\cdot(\vec{r}_{j}-\vec{w}t)}\right)\delta(\omega-\vec{w}\cdot\vec{k})\nonumber\\
&&=-i(8\pi^2e)\frac{[\vec{k}-(\omega\vec{w}/c^2)]}
{\kappa^2}\left(\sum_{j=1}^{N}e^{-i\kappa\cdot\xi_j}\right)\delta(\omega-\vec{w}\cdot\vec{k}),\label{cov-due}
\end{eqnarray}
where the $4$-vector $\xi_j=(\vec{r}_j,ct)$ $(j=1,..,N)$ represents
the space-time coordinates of the $N$ electrons in the laboratory
reference frame.

The covariance of the electric field of the $N$ electron bunch - see
Eq.(\ref{cov-due}) or equivalently Eq.(\ref{cov-bisbis}) - can be
finally verified, as in following argued. In fact, looking at the
integrand of Eq.(\ref{cov-uno}) and at Eq.(\ref{cov-due}), the term
$[k_z-(\omega w/c^2)]\delta(\omega-\vec{w}\cdot\vec{k})$ is the only
non-Lorentz-invariant quantity whose covariance has to be checked.
Under a Lorentz transformation from the laboratory to the rest
reference frame ($\mathbb{R}\Longrightarrow\mathbb{R'}$), such a
quantity transforms indeed as, see also \ref{lorentz} and
\ref{vq-lab}:
\begin{displaymath}
\left\{
\begin{array}{l}
\delta(\omega-\vec{w}\cdot\vec{k})\longrightarrow\delta[\gamma(\omega'+wk'_z)-w\gamma(k'_z+\beta\omega'/c))]
=\delta(\frac{\omega'}{\gamma})=\gamma\delta(\omega')\\
(k_z-\omega
w/c^2)\longrightarrow\gamma(k'_z+\beta\omega'/c)-\gamma(\omega'+\beta
ck'_z)w/c^2= \frac{k'_z}{\gamma}
\end{array} \right.
\end{displaymath}
Finally, under a Lorentz transformation
$\mathbb{R}\Longrightarrow\mathbb{R'}$,
Eqs.(\ref{cov-uno},\ref{cov-due}) transform as:
\begin{eqnarray}
\vec{E}(\vec{r},t)&&=-\frac{iec}{2\pi^2}\int d^4\kappa
e^{i\kappa\cdot\xi}\frac{\sum_{j=1}^{N}e^{-i\kappa\cdot\xi_j}}{\kappa^2}
[\vec{k}-(\omega\vec{w}/c^2)]\delta(\omega-\vec{w}\cdot\vec{k})=\nonumber\\
&&=-\frac{iec}{2\pi^2}\int
d^4\kappa'e^{i\kappa'\cdot\xi'}\frac{\sum_{j=1}^{N}e^{-i\kappa'\cdot\xi'_j}}{\kappa'^2}
\left( \begin{array}{c}
\gamma k'_x\\
\gamma k'_y\\
k'_z\\
\end{array}\right)\delta(\omega')=\nonumber\\
&&=\left( \begin{array}{c}
\gamma E'_x(\vec{r'},t') \\
\gamma E'_y(\vec{r'},t') \\
E'_z(\vec{r'},t')\\
\end{array}\right),\label{cov-tre}
\end{eqnarray}
where, in the last term of Eq.(\ref{cov-tre}), the three components
of the charge electric field $\vec{E'}(\vec{r'},t')$ in the rest
reference frame $\mathbb{R'}$ can be identified, see also
\ref{cov-rest}:
\begin{eqnarray}
\vec{E'}(\vec{r'},t')=-\frac{i e c}{2\pi^2}\int
d^4\kappa'e^{i\kappa'\cdot\xi'}\frac{\sum_{j=1}^{N}e^{-i\kappa'\cdot\xi'_j}}{\kappa'^2}\vec{k'}\delta(\omega').
\label{cov-cinque}
\end{eqnarray}
Equation(\ref{cov-cinque}) is the explicitly covariant Fourier
representation of the electric field of a $N$ electron bunch in the
rest reference frame $\mathbb{R'}$, where $(x',y',z',ct')$ and
$(k'_x,k'_y,k'_z,\omega'/c)$ are, respectively, the space-time and
the conjugate Fourier wavevector-frequency $4$-vectors of
$\mathbb{R'}$ and $\xi'_j=(\vec{r'}_j,ct')$ $(j=1,..,N)$ is the
space-time $4$-vectors representing the $N$ electron coordinates in
$\mathbb{R'}$, see also \ref{cov-rest} and \ref{vq-lab}.

In conclusion, the expression of the {\it virtual quanta} field, see
Eq.(\ref{cov-due}) or equivalently Eq.(\ref{cov-bisbis}), which in
the present and other works
\cite{gian,gianbis,gianter,gianquater,gian5,gian6} is the starting
point to the formulation of the radiation energy spectrum of a $N$
electron bunch, is demonstrated to meet the covariant constraint.
The formulation of the radiation energy spectrum - see
Eqs.(\ref{tre},\ref{quattro1},\ref{cento-uno}) - is also
demonstrated to be covariant and temporal causality consistent.
Equations (\ref{tre},\ref{quattro1},\ref{cento-uno}) are indeed the
final result of a series of temporal-causality-consistent formal
steps which preserve and transmit the original covariance of the
{\it virtual quanta} field of the $N$ electron beam into the
radiation energy spectrum.

Finally, about the invariance of the transverse projection of the
$N$ electron density with respect to the direction of motion and
about the covariant imprinting of it on the charge electric field
and, consequently, on the radiation field and on the radiation
energy spectrum, the following observation can be done. The effect
of the transverse electron coordinates
$\vec{\rho}_{0j}=(x_{0j},y_{0j})$ $(j=1,..,N)$ on the Fourier
transformation of the charge electric field is driven by a phase
factor, see Eqs.(\ref{cov-uno},\ref{cov-due}) and
Eqs.(\ref{cov-tre},\ref{cov-cinque}). This is a function of the
Lorentz invariant quantity $k_xx_{0j}+k_yy_{0j}$ $(j=1,..,N)$, where
$(k_x,k_y)$ are the transverse component of the wave-vector. With
reference to Eqs.(\ref{cov-uno},\ref{cov-due}) and
Eqs.(\ref{cov-tre},\ref{cov-cinque}), the charge electric field
shows a dependence on the transverse electron coordinates which,
under a Lorentz transformation from the laboratory to the rest
reference frame, transfers and transforms in a covariant way and
whose observability at a given wavelength remains unchanged, this
being determined by the Lorentz-invariant scalar product of the
electron transverse coordinates and the transverse components of the
wave-vector. In the light of such a meaning, the statement that, see
also end of section \ref{pseudo-virtual}, the Lorentz invariance of
the transverse projection of the $N$ electron density is expected to
leave a covariant mark on the $N$ electron radiation field
amplitudes - Eqs.(\ref{tre},\ref{quattro1}) - and, consequently, on
both the temporal coherent and incoherent components of the
radiation energy spectrum - see Eq.(\ref{cento-uno}) - has to be
read and integrated with the statement that the observability of the
covariant effect of the transverse coordinates on the charge
electric field remains unchanged, passing from an inertial reference
frame to the other, this being a function of a Lorentz-invariant
quantity.

\section{Conclusions}

In the present paper, the issue of the covariant formulation of the
radiation energy spectrum and, consequently, of the charge form
factor of a $N$ electron bunch is argued. The $N$ electron bunch is
supposed to collide, at a normal angle of incidence, onto a metallic
screen $S$ which has an arbitrary shape and size (either finite
$S<\infty$ or infinite $S=\infty$) and is supposed to behave as an
ideal conductor in the wavelength region of interest. As the
covariance of a charged distribution is expected to evolve from a
charge-point-like into a charge-density-like one when passing from a
single electron to a $N$ electron bunch, the covariance of the
radiation energy spectrum as well is expected to behave in the same
way when a $N$ electron bunch is considered instead of a single
electron. In order to verify the covariant formulation of the
radiation energy spectrum, the formal steps leading from the virtual
quanta field to the transition radiation energy spectrum of a $N$
electron bunch have been explicitly derived and checked to meet,
first, the temporal causality constraint and, then, to be covariance
consistent. Finally, the expression of the {\it virtual quanta}
field, which, in the present paper and in other papers
\cite{gian,gianbis,gianter,gianquater,gian5,gian6}, is the starting
point to achieve the formula of the radiation energy spectrum, has
been checked to transform in a covariant way under a Lorentz
transformation from the laboratory reference frame ($\mathbb{R}$) to
the reference frame of rest of the $N$ electron bunch
($\mathbb{R'}$). In order to perform such a covariance check, the
Fourier representations of fields and charged distributions have
been suitably extended from the ordinary $3$-dimensional space of
the spatial coordinates and the conjugate Fourier wave-vectors into
the Lorentz $4$-spaces of the space-time and of the conjugate
Fourier wavevector-frequency in both $\mathbb{R}$ and $\mathbb{R'}$.
In a temporal causal and covariant formulation of the transition
radiation energy spectrum of a $N$ electron bunch, the invariance of
the projection of the electron density in the transverse plane with
respect to the direction of motion manifests itself as a covariant
feature of the $N$ single electron amplitude composing the radiation
field and, consequently, of both the temporal coherent and
incoherent components of the transition radiation energy spectrum.

%\section{ACKNOWLEDGMENTS}

%\section{}
%\label{}

% \appendix

% \section{}
% \label{}

\appendix
\section{Lorentz transformations of coordinates and fields}\label{lorentz}

In the present work, the case of a bunch of $N$ electrons in a
rectilinear and uniform motion with a common velocity
$\vec{w}=(0,0,w)$ along the positive direction of the z-axis of the
laboratory reference frame $\mathbb{R}$ is considered. In the
present paper, Lorentz transformations from the laboratory reference
frame $\mathbb{R}$ to the reference frame of rest $\mathbb{R'}$ of
the $N$ electron bunch - and viceversa - are applied to fields and
spatial distributions of charges. Therefore, for the sake of easy
reading, the main results are below reported.

The $4$-vectors of $\mathbb{R}$, the space-time $(x,y,z,ct)$ and the
conjugate Fourier wavevector-frequency $(k_x,k_y,k_z,\omega/c)$,
Lorentz-transform into the respective counterparts $(x',y',z',ct')$
and $(k'_x,k'_y,k'_z,\omega'/c)$ of $\mathbb{R'}$ according to
\cite{jack} ($\mathbb{R'}\Longrightarrow\mathbb{R}$):
\begin{displaymath}
\left\{
\begin{array}{l}
x'=x\\
y'=y\\
z'=\gamma (z-\beta x_0)\\
x'_0=\gamma (x_0- \beta z)
 \end{array} \right.\label{poppi}
\end{displaymath}
\begin{displaymath}
\left\{
\begin{array}{l}
k'_x=k_x\\
k'_y=k_y\\
k'_z=\gamma (k_z-\beta k_0)\\
k'_0=\gamma (k_0- \beta k_z)
 \end{array} \right.
\end{displaymath}
where $\beta=w/c$ and $\gamma=\frac{1}{\sqrt{1-\beta^2}}$.

As for the electromagnetic fields $\vec{E}$ and $\vec{B}$, Lorentz
transformations $\mathbb{R'}\Longrightarrow\mathbb{R}$ read:
\begin{displaymath}
\left\{
\begin{array}{l}
E'_x(\vec{r'},t')=\gamma(E_x(\vec{r},t)-\beta B_y(\vec{r},t))\\
E'_y(\vec{r'},t')=\gamma(E_y(\vec{r},t)+\beta B_x(\vec{r},t))\\
E'_z(\vec{r'},t')=E_z(\vec{r},t)
 \end{array} \right.
\end{displaymath}
and
\begin{displaymath}
\left\{
\begin{array}{l}
B'_x(\vec{r'},t')=\gamma(B_x(\vec{r},t)+\beta E_y(\vec{r},t))\\
B'_y(\vec{r'},t')=\gamma(B_y(\vec{r},t)-\beta E_x(\vec{r},t))\\
B'_z(\vec{r'},t')=B_z(\vec{r},t)
 \end{array} \right.
\end{displaymath}
The inverse Lorentz transformations
$\mathbb{R}\Longrightarrow\mathbb{R'}$ for coordinates and fields
can be obtained from previous equations by substituting
$\beta\rightarrow-\beta$ and by exchanging the prime in field
components and coordinates.

\section{Covariance of the spatial density of a bunch of $N$ electrons}\label{cov}

$N$ electrons in a bunch are supposed to move with a common
rectilinear and uniform velocity in the laboratory reference frame
$\mathbb{R}$. In the frame of reference $\mathbb{R'}$ where each
electron is at rest, the $N$ electrons can be described in terms of
a ``static" distribution function of the particle density:
\begin{equation}
\rho'(\vec{r'},t')=\rho'(\vec{r'},0)=\sum_{j=1}^N\delta(\vec{r'}-
\vec{r'}_j) \label{cov1}
\end{equation}
where $\vec{r'}=(x',y',z')$ and $\vec{r'}_j=(x'_j,y'_j,z'_j)$
$(j=1,..,N)$ are the $N$ electron coordinates in the rest reference
frame $\mathbb{R'}$. In the conjugate Fourier space
$(k'_x,k'_y,k'_z)$ of the ordinary $3$-space of the spatial
coordinates $(x',y',z')$ of $\mathbb{R'}$, the equation above reads
\begin{equation}
\rho'(\vec{r',t'})=\frac{1}{(2\pi)^3}\int
d\vec{k'}e^{i\vec{k'}\cdot\vec{r'}}\rho'(\vec{k'})\label{cov2}
\end{equation}
where
\begin{equation}
\rho'(\vec{k'})=\sum_{j=1}^{N}e^{-i\vec{k'}\cdot
\vec{r'_j}}=\sum_{j=1}^{N}e^{-i(k'_xx'_j+k'_yy'_j+k'_zz'_j)}\label{cov3}
\end{equation}
is the Fourier transform of the distribution function of the $N$
particle density. Since the bunch density - as represented by
Eq.(\ref{cov2}) - is only defined in a subspace $(k'_x,k'_y,k'_z)$
of the wavevector-frequency 4-vector $(k'_x,k'_y,k'_z,\omega'/c)$ of
$\mathbb{R'}$, it cannot explicitly show the expected
Lorentz-covariance. The explicit covariance of the spatial density
of the $N$ electrons can be directly retrieved by extending the
Fourier representation of Eqs.(\ref{cov2},\ref{cov3}) into the
wavevector-frequency $(k'_x,k'_y,k'_z,\omega'/c)$ and space-time
$(x',y',z',ct')$ $4$-spaces of the rest reference frame
$\mathbb{R'}$. Such a result can be achieved by extending the
integral of Eq.(\ref{cov2}) into the $4$-space
$\kappa=(k'_x,k'_y,k'_z,\omega'/c)$ and the integrand into itself
times $\delta(\omega')e^{i\omega't'}e^{-i\omega't'}$ where
$\delta(\omega')$ is a delta Dirac function. Following such a formal
procedure, Eq.(\ref{cov2}) can be reformulated as follows:
\begin{eqnarray}
\rho'(\vec{r',t'})&&=\frac{c}{(2\pi)^3}\int
d\vec{k'}d(\omega'/c)e^{i(\vec{k'}\cdot\vec{r'}-\omega'
t')}\left(\sum_{j=1}^{N}e^{-i(\vec{k'}\cdot \vec{r'_j}-\omega'
t')}\right)\delta(\omega')=\nonumber\\
&&=\frac{c}{(2\pi)^3}\int
d^4\kappa'e^{i\kappa'\cdot\xi'}\left(\sum_{j=1}^{N}e^{-i\kappa'\cdot\xi'_j}\right)\delta(\omega')=\nonumber\\
&&=\frac{c}{(2\pi)^4}\int
d^4\kappa'e^{i\kappa'\cdot\xi'}\rho(\vec{k'},\omega')\label{cov4}
\end{eqnarray}
where
\begin{equation}
\rho'(\vec{k'},\omega')=2\pi\left(\sum_{j=1}^{N}e^{-i\kappa'\cdot\xi'_j}\right)\delta(\omega')\label{cov5}
\end{equation}
with $\kappa'=(k'_x,k'_y,k'_z,\omega'/c)$ and $\xi'=(x',y',z',ct')$
and where $\xi'_j=(x'_j,y'_j,z'_j,ct')$ $(j=1,..,N)$ are the
space-time coordinate $4$-vectors of the electrons in the rest
reference frame $\mathbb{R'}$. Thanks to the extension of the
Fourier representation of the spatial density of the $N$ electron
bunch into the $4$-space of the rest reference frame $\mathbb{R'}$ -
see Eqs.(\ref{cov4},\ref{cov5}) - the expected covariance of the
charge density can be explicitly recovered. The following quantities
are indeed Lorentz-invariant
\begin{displaymath}
\left\{
\begin{array}{l}
d^4\kappa'=d\vec{k'}d(\omega'/c)=d^4\kappa\\
\kappa'\cdot\xi'=k'_xx'+k'_yy'+k'_zz'-\omega't'=\kappa\cdot\xi\\
\kappa'\cdot\xi'_j=k'_xx'_j+k'_yy'_j+k'_zz'_j-\omega't'=\kappa\cdot\xi_j
\end{array} \right.
\end{displaymath}
Consequently, under a Lorentz transformation from the rest to the
laboratory reference frame ($\mathbb{R'}\Longrightarrow\mathbb{R}$),
Eq.(\ref{cov5}) transforms as follows, see also \ref{lorentz}:
\begin{eqnarray}
\rho'(\vec{k'},\omega')&&=2\pi\left(\sum_{j=1}^{N}e^{-i\kappa'\cdot\xi'_j}\right)\delta(\omega')
=2\pi\left(\sum_{j=1}^{N}e^{-i\kappa\cdot\xi_j}\right)\delta[\gamma
(\omega-wk_z)]=\nonumber\\
&&=2\pi\left(\sum_{j=1}^{N}e^{-i\kappa\cdot\xi_j}\right)\frac{\delta(\omega-\vec{w}\cdot\vec{k})}{\gamma}=
\frac{1}{\gamma}\rho(\vec{k},\omega),\label{cov6}
\end{eqnarray}
where
\begin{eqnarray}
\rho(\vec{k},\omega)=2\pi\left(\sum_{j=1}^{N}e^{-i\kappa\cdot\xi_j}\right)\delta(\omega-\vec{w}\cdot\vec{k})
\label{cov6-bis}
\end{eqnarray}
represents the $4$-dimensional Fourier transformation of the
distribution function of the spatial density of the $N$ electron
bunch in the laboratory reference frame $\mathbb{R}$. In the
equation above, $\xi_j=(x_j,y_j,z_j,ct)$ $(j=1,..,N)$ are the
$4$-vector space-time coordinates of the electrons in $\mathbb{R}$
as well as $\xi=(x,y,z,ct)$ and $\kappa=(k_x,k_y,k_z,\omega/c)$ are,
respectively, the space-time and the conjugate Fourier
wavevector-frequency Lorentz $4$-vectors of $\mathbb{R}$.

In conclusion, under a Lorentz transformation
$\mathbb{R'}\Longrightarrow\mathbb{R}$, the distribution function of
the spatial density of the $N$ electrons - see
Eqs.(\ref{cov1},\ref{cov2}) and Eqs.(\ref{cov6},\ref{cov6-bis}) -
transforms according to the expected covariance:
\begin{equation}
\rho'(\vec{r'},t')=\frac{1}{\gamma}\rho(\vec{r},t),\label{cov7}
\end{equation}
where
\begin{eqnarray}
\rho(\vec{r},t)&&=\frac{c}{(2\pi)^3}\int
d^4\kappa\,e^{i\kappa\cdot\xi}\left(\sum_{j=1}^{N}e^{-i\kappa\cdot\xi_j}\right)\delta(\omega-
\vec{w}\cdot\vec{k}) =\nonumber\\
&&=\frac{1}{(2\pi)^3}\int d\vec{k}\,
e^{i\vec{k}\cdot(\vec{r}-\vec{w}t)}\left(\sum_{j=1}^{N}e^{-i\vec{k}\cdot
(\vec{r_j}-\vec{w}t)}\right)=\rho(\vec{r}-\vec{w}t)=\nonumber\\
&&=\frac{1}{(2\pi)^3}\int d\vec{k}\,
e^{i\vec{k}\cdot\vec{r}_{0}}\left(\sum_{j=1}^{N}e^{-i\vec{k}\cdot
\vec{r}_{0j}}\right)=\sum_{j=1}^N\delta(\vec{r}_{0}-
\vec{r}_{0j})=\rho(\vec{r}_{0},0)\label{cov8}
\end{eqnarray}
represents the distribution function of the $N$ electron density in
the laboratory reference frame $\mathbb{R}$, see also Eqs.($5$,$6$)
in \cite{gianbis}.

The distribution function of the spatial density of the $N$
electrons - as represented by Eq.(\ref{cov8}) - is invariant under a
time-space translation in the laboratory reference frame
$\mathbb{R}$
\begin{displaymath}
\left\{
\begin{array}{l}
\vec{r}=\vec{r}_0+\vec{w}t\\
\vec{r}_j=\vec{r}_{0j}+\vec{w}t
 \end{array} \right.
\end{displaymath}
or, in other words, it is ``frozen-in-time" as already mentioned in
the present paper where the case of a $N$ electron bunch in a
rectilinear and uniform motion is considered. Looking at both
integrand of Eqs.(\ref{cov4},\ref{cov5},\ref{cov6-bis},\ref{cov8}),
the presence of a delta Dirac function and, in particular, the
argument of this delta are instructive about the attribute given to
the charge distribution function to be ``static" in the rest
reference frame $\mathbb{R'}$ or ``frozen-in-time" in the laboratory
reference frame $\mathbb{R}$. In the reference frame $\mathbb{R'}$,
where the $N$ electrons are at rest, $\delta(\omega')$ in
Eq.(\ref{cov5}) is indicating that only a wave with $\omega'=0$
constitutes the relevant harmonic contribution to the Fourier
transformation of the charge density. Conversely, in the laboratory
reference frame $\mathbb{R}$, where all the $N$ electrons move with
a common rectilinear and uniform velocity $\vec{w}$, the delta Dirac
$\delta(\omega- \vec{w}\cdot\vec{k})$ in Eq.(\ref{cov8}) is
indicating that a plane wave traveling with the same velocity
$\vec{w}$ as the electron bunch constitutes the only relevant
harmonic contribution to the Fourier transformation of the charge
density.

\section{Electric field of a $N$ electron bunch in the rest reference frame}\label{cov-rest}

In the rest reference frame $\mathbb{R'}$ of the $N$ electron bunch,
the magnetic field $\vec{B'}(\vec{r'},t')\equiv 0$. The electric
field $\vec{E'}(\vec{r'})=-\vec{\nabla}\Phi'(\vec{r'})$ can be
obtained from the Poisson equation for the scalar potential $\Phi$:
\begin{equation}
-\nabla^2\Phi'(\vec{r'})=4\pi e\,\rho'(\vec{r'}).\label{rest1}
\end{equation}
In the Fourier $3$-space $\vec{k'}=(k'_x,k'_y,k'_z)$ of
$\mathbb{R'}$, where the scalar potential and the charge density
read
\begin{displaymath}
\left\{
\begin{array}{l}
\Phi'(\vec{r'})=\frac{1}{(2\pi)^3}\int d\vec{k'}e^{i\vec{k'}\cdot\vec{r'}}\Phi'(\vec{k'})\\
\rho'(\vec{r'})=\frac{1}{(2\pi)^3}\int
d\vec{k'}e^{i\vec{k'}\cdot\vec{r'}}\rho'(\vec{k'})
 \end{array} \right.
\end{displaymath}
the Poisson equation reads
\begin{equation}
\Phi'(\vec{k'})=4\pi e\,\frac{\rho'(\vec{k'})}{k'^2}.\label{rest2}
\end{equation}
and the electric field reads
\begin{eqnarray}
\vec{E'}(\vec{r'})&&=-\vec{\nabla}\Phi'(\vec{r'})=-\frac{i\,e}{2\pi^2}\int
d\vec{k'}e^{i\vec{k'}\cdot\vec{r'}}\frac{\vec{k'}}{k'^2}\rho'(\vec{k'})=\nonumber\\
&&=-\frac{i\,e}{2\pi^2}\int
d\vec{k'}e^{i\vec{k'}\cdot\vec{r'}}\frac{\vec{k'}}{(k'^2_x+k'^2_y+k'^2_z)}\left(\sum_{j=1}^{N}e^{-i\vec{k'}\cdot
\vec{r'_j}}\right).\label{rest4}
\end{eqnarray}
where the explicit expression of $\rho'(\vec{k'})$ is given in
Eq.(\ref{cov3}).

The Fourier representation of the electric field of a $N$ electron
bunch in the rest reference frame $\mathbb{R'}$  - as given in
Eq.(\ref{rest4}) - is evidently non-explicitly covariant being
restricted to the subspaces $\vec{k'}=(k'_x,k'_y,k'_z)$ and
$\vec{r'}=(r'_x,r'_y,r'_z)$, respectively, of the
wavevector-frequency $\vec{\kappa'}=(k'_x,k'_y,k'_z,\omega'/c)$ and
space-time $\vec{\xi'}=(x',y',z',ct')$ Lorentz $4$-vectors of
$\mathbb{R'}$. The covariance of the electric field can be
explicitly retrieved by suitably extending the Fourier
transformation in Eq.(\ref{rest4}) into the Fourier $4$-space
$(k'_x,k'_y,k'_z,\omega'/c)$. Following the same formal procedure
already described in \ref{cov} for the charge density, the Fourier
transformation of the electric field of the $N$ electron bunch in
the rest reference frame $\mathbb{R'}$ - Eq.(\ref{rest4}) - can be
reformulated as:
\begin{eqnarray}
\vec{E'}(\vec{r'})&&= -\frac{ie}{2\pi^2}\int
d\vec{k'}e^{i\vec{k'}\cdot\vec{r'}}\frac{\vec{k'}}{(k'^2_x+k'^2_y+k'^2_z)}(\sum_{j=1}^{N}e^{-i\vec{k'}\cdot
\vec{r'_j}})=\nonumber\\
&&=-\frac{i\,e}{2\pi^2}\int
d\vec{k'}d\omega'e^{i(\vec{k'}\cdot\vec{r'}-\omega't')}(\sum_{j=1}^{N}e^{-i(\vec{k'}\cdot
\vec{r'_j}-\omega't')})\frac{\vec{k'}\delta(\omega')}{[k'^2_x+k'^2_y+k'^2_z-(\omega'/c)^2]}=\nonumber\\
&&=-\frac{iec}{2\pi^2}\int d^4\kappa'e^{i\kappa'\cdot\xi'}
\frac{(\sum_{j=1}^{N}e^{-i\kappa'\cdot\xi'_j})}{\kappa'^2}\vec{k'}\delta(\omega')=\nonumber\\
&&=\frac{c}{(2\pi)^4}\int d^4\kappa' \,
e^{i\kappa'\cdot\xi'}\vec{E'}(\vec{k'},\omega'),\label{rest5}
\end{eqnarray}
where $\xi'_j=(\vec{r'}_j,ct')=(x'_j,y'_j,z'_j,ct')$ are the
space-time coordinates of the $N$ electrons $(j=1,..,N)$ in
$\mathbb{R'}$. In the integrand of the equation above, the only
non-Lorentz invariant quantity is the Fourier transform of the
electric field $\vec{E'}(\vec{k'},\omega')$:
\begin{eqnarray}
\vec{E'}(\vec{k'},\omega')=-i(8\pi^2e)\frac{(\sum_{j=1}^{N}e^{-i\kappa'\cdot
\xi'_j})}{\kappa'^2}\vec{k'}\delta(\omega').\label{rest6}
\end{eqnarray}
The term $\vec{k'}\delta(\omega')$ is the only quantity in
Eqs.(\ref{rest5},\ref{rest6}) whose covariance has to be checked.
Under a Lorentz transformation from the rest to the laboratory
reference frame $(\mathbb{R'}\Longrightarrow\mathbb{R})$ - see also
\ref{lorentz} - such a quantity transforms as:
\begin{displaymath}
\left\{
\begin{array}{l}
k'_{x,y}\delta(\omega')\longrightarrow k_{x,y}\delta[\gamma(\omega-wk_z)]=
\frac{k_{x,y}}{\gamma}\delta(\omega-wk_z)\\
k'_z\delta(\omega')\longrightarrow
\gamma(k_z-\frac{w\omega}{c^2})\delta[\gamma(\omega-wk_z)]=(k_z-\frac{w\omega}{c^2})\delta(\omega-wk_z)
 \end{array} \right.
\end{displaymath}
where $\vec{w}=(0,0,w)$ is the velocity of the electrons in the
laboratory reference frame $\mathbb{R}$. Finally, under a Lorentz
transformation $\mathbb{R'}\Longrightarrow\mathbb{R}$, the electric
field of a $N$ electron bunch in the rest reference frame
$\mathbb{R'}$ - see Eqs.(\ref{rest4},\ref{rest5},\ref{rest6}) -
transforms into
\begin{eqnarray}
\vec{E'}(\vec{r'})&&=\frac{c}{(2\pi)^4}\int d^4\kappa' \,
e^{i\kappa'\cdot\xi'}\vec{E'}(\vec{k'},\omega')=\nonumber\\
&&=-\frac{i e c}{2\pi^2}\int d^4\kappa \,
e^{i\kappa\cdot\xi}\frac{(\sum_{j=1}^{N}e^{-i\kappa\cdot
\xi_j})}{\kappa^2} \left( \begin{array}{c}
k_x/\gamma \\
k_y/\gamma \\
k_z-w \omega/c^2\\
\end{array}\right)\delta(\omega-\vec{w}\cdot\vec{k})=\nonumber\\
&&=\left( \begin{array}{c}
E_x(\vec{r},t)/\gamma \\
E_y(\vec{r},t)/\gamma \\
E_z(\vec{r},t)\\
\end{array}\right)\label{rest7}
\end{eqnarray}
where, in the equation above, the $3$ components of the electric
field of the $N$ electron bunch in the laboratory reference frame
$\mathbb{R}$ can be recognized, see
Eqs.(\ref{cov-uno},\ref{cov-due}) and also \ref{vq-lab}:
\begin{eqnarray}
\vec{E}(\vec{r},t)=-\frac{i e c}{2\pi^2}\int d^4\kappa \,
e^{i\kappa\cdot\xi}\frac{(\vec{k}-\vec{w}\omega/c^2)}{\kappa^2}\left(\sum_{j=1}^{N}e^{-i\kappa\cdot
\xi_j}\right)\delta(\omega-\vec{w}\cdot\vec{k}).\label{rest8}
\end{eqnarray}

\section{Electromagnetic field of a $N$ electron bunch in the laboratory reference frame}\label{vq-lab}

In the case of a bunch of $N$ electrons in motion in vacuum with a
common uniform and rectilinear velocity $\vec{w}$ in the laboratory
reference frame $\mathbb{R}$, the Fourier representation of the
propagation equations of the $4$-potential $(\vec{A},\Phi)$ in the
$4$-space $(\vec{k},\omega/c)$ of $\mathbb{R}$ reads in the gauge of
Lorentz as
\begin{displaymath}
\left\{
\begin{array}{l}
(-\frac{\omega^2}{c^2}+k^2)\vec{A}(\vec{k},\omega)=\frac{4 \pi\,e}{c}\vec{w}\rho(\vec{k},\omega)\\
(-\frac{\omega^2}{c^2}+k^2)\Phi(\vec{k},\omega)=4\pi
e\,\rho(\vec{k},\omega)
 \end{array} \right.
\end{displaymath}
where $\vec{J}(\vec{k},\omega)=e\rho(\vec{k},\omega)\vec{w}$ is the
Fourier transformation of the current density vector. With reference
to the equations above, the Fourier transformation of the charge
electric field reads:
\begin{eqnarray}
\vec{E}(\vec{k},\omega)&&=-i\vec{k}\Phi(\vec{k},\omega)+
\frac{i\omega}{c}\vec{A}(\vec{k},\omega)=\nonumber\\
&&=-i4\pi e\,\frac{(\vec{k}-\omega\vec{w}/c^2)}
{[k^2-(\omega/c)^2]}\rho(\vec{k},\omega).\label{vq-lab-uno}
\end{eqnarray}
With reference to Eq.(\ref{cov6-bis}) in \ref{cov}, the Fourier
representation of the electric field of the $N$ electron bunch - as
given in Eq.(\ref{vq-lab-uno}) - can be expressed in an explicitly
covariant form in the space-time $\xi=(\vec{r},ct)$ and in the
$4$-space of the conjugate Fourier wavevector-frequency
$\vec{\kappa}=(\vec{k},\omega/c)$ of the laboratory reference frame
$\mathbb{R}$ , see also Eqs.(\ref{cov-uno},\ref{cov-due}) and
Eq.(\ref{rest8}) in \ref{cov-rest}:
\begin{eqnarray}
\vec{E}(\vec{k},\omega)=-i8\pi^2e\frac{(\vec{k}-\omega\vec{w}/c^2)}
{\kappa^2}\left(\sum_{j=1}^{N}e^{-i\kappa\cdot\xi_j}\right)\delta(\omega-\vec{w}\cdot\vec{k}),\label{vq-lab-due}
\end{eqnarray}
where $\kappa^2=k^2-(\omega/c)^2$ and $\xi_j=(\vec{r}_j,ct)$
$(j=1,..,N)$ are the space-time $4$-vectors of the $N$ electron
coordinates in the laboratory reference frame $\mathbb{R}$.

Under a Lorentz transformation from the laboratory to the rest
reference frame $(\mathbb{R}\Longrightarrow\mathbb{R'})$, the
non-Lorentz-invariant quantities in Eq.(\ref{vq-lab-due}) transforms
as follows:
\begin{displaymath}
\left\{
\begin{array}{l}
k_z-\omega w/c^2=k'_z/\gamma
\\
\delta(\omega-wk_z)=\gamma\delta(\omega')
\end{array} \right.
\end{displaymath}
see also \ref{lorentz}. Taking into account the equations above, in
the laboratory reference frame $\mathbb{R}$, the covariance of the
$4$-dimensional Fourier representation of the electric field of the
$N$ electron bunch - Eq.(\ref{vq-lab-due}) - can be finally checked.
Under a Lorentz transformation
$\mathbb{R}\Longrightarrow\mathbb{R'}$, Eq.(\ref{vq-lab-due})
transforms indeed as
\begin{eqnarray}
\vec{E}(\vec{k},\omega)&&=-i8\pi^2e\frac{\left(\sum_{j=1}^{N}e^{-i\kappa'\cdot\xi'_j}\right)}
{\kappa'^2}\left(
\begin{array}{c}
\gamma k'_x \\
\gamma k'_y \\
k'_z\\
\end{array}\right)\delta(\omega')=\left( \begin{array}{c}
\gamma E'_x(\vec{k'},\omega')\\
\gamma E'_y(\vec{k'},\omega') \\
E'_z(\vec{k'},\omega')\\
\end{array}\right)\label{vq-lab-duecento}
\end{eqnarray}
where, in the second term of previous equation, the explicitly
covariant Fourier representation of the charge electric field
$\vec{E'}(\vec{k'},\omega')$ in the rest reference frame
$\mathbb{R'}$ can be recognized, see Eq.(\ref{cov-cinque}) and also
Eq.(\ref{rest6}) in \ref{cov-rest}.

For the sake of completeness, the electric field of $N$ electrons in
the rest reference frame $\mathbb{R'}$ - see second term of
Eq.(\ref{vq-lab-duecento}) or, equivalently, Eq.(\ref{cov-cinque})
or Eq.(\ref{rest6}) in \ref{cov-rest} - will be in the following
obtained by Lorentz transforming the electric and magnetic fields in
the laboratory frame of reference $\mathbb{R}$. In the laboratory
reference frame $\mathbb{R}$, the Fourier transformation of the
magnetic field by a $N$ electron bunch reads:
\begin{eqnarray}
\vec{B}(\vec{k},\omega)=i\vec{k}\times\vec{A}(\vec{k},\omega)=\frac{i8\pi^2e(\sum_{j=1}^{N}e^{-i\kappa\cdot
\xi_j})}{c\kappa^2}\left( \begin{array}{c}
wk_y \\
-wk_x \\
0\\
\end{array}\right)\delta(\omega-\vec{k}\cdot\vec{w}).\label{vq-lab-quattro}
\end{eqnarray}
The $4$-dimensional Fourier representation of the transverse
components of the charge electric field in $\mathbb{R'}$ can be
obtained from the corresponding components of the electromagnetic
field in $\mathbb{R}$ as (see also \ref{lorentz}):
\begin{eqnarray}
E'_{(x,y)}(\vec{k'},\omega')&&=\gamma[E_{(x,y)}(\vec{k},\omega)\mp\beta
B_{(y,x)}(\vec{k},\omega)]=\nonumber\\
&&=\gamma(-i8\pi^2
e)\frac{\left(\sum_{j=1}^{N}e^{-i\kappa\cdot\xi_j}\right)}{\kappa^2}\delta(\omega-\vec{w}\cdot\vec{k})[\left(
\begin{array}{c}
k_x \\
k_y \\
\end{array}\right)\pm\beta^2\left(
\begin{array}{c}
-k_x \\
k_y \\
\end{array}\right)]=\nonumber\\
&&=\frac{-i8\pi^2e}{\gamma}
\frac{\left(\sum_{j=1}^{N}e^{-i\kappa\cdot\xi_j}\right)}{\kappa^2}\delta(\omega-wk_z)\left(
\begin{array}{c}
k_x \\
k_y \\
\end{array}\right)=\nonumber\\
&&=-i8\pi^2e\frac{\left(\sum_{j=1}^{N}e^{-i\kappa'\cdot\xi'_j}\right)}{\kappa'^2}\delta(\omega')\left(
\begin{array}{c}
k'_x \\
k'_y \\
\end{array}\right)\label{vq-lab-trecento}
\end{eqnarray}
where in the last term of previous equation the transverse component
of the Fourier transform of the charge electric field in the rest
reference frame $\mathbb{R'}$ can be recognized (see \ref{cov-rest},
Eq.(\ref{rest6}) in particular). Concerning the longitudinal
component $E_{z}(\vec{k},\omega)$, the corresponding covariant
transformation is trivial, see Eq.(\ref{vq-lab-duecento}) or
Eq.(\ref{rest6}).

%\end{linenumbers}

\end{document}